\documentclass[pra,10pt,final,twocolumn,showpacs,superscriptaddress,aps]{revtex4-1}
\usepackage{graphicx}  
\usepackage{amssymb}   
\usepackage{amsmath}   
\usepackage{color}     
\usepackage{dsfont}
\usepackage{ulem}
\usepackage{physics}
\usepackage{dsfont}
\usepackage{romannum}
\usepackage{subcaption}
\usepackage[justification=raggedright,labelfont={bf}]{caption}

\begin{document}

\title{Optical solitons in near $\mathcal{PT}-$symmetric Rosen-Morse potential}

\author{K. Hari}
\affiliation{Department of Physics, National Institute of Technology, Tiruchirappalli - 620015, Tamil Nadu, India.}

\author{K. Manikandan}
\email{manikandan.cnld@gmail.com}
\affiliation{Department of Physics, National Institute of Technology, Tiruchirappalli - 620015, Tamil Nadu, India.}

\author{R. Sankaranarayanan}
 \email{sankar@nitt.edu}
\affiliation{Department of Physics, National Institute of Technology, Tiruchirappalli - 620015, Tamil Nadu, India.}

\begin{abstract}
We investigate the existence of stable soliton solution in a system described by complex Ginzburg-Landau (CGL) equation with near parity reflection - time reversal ($\mathcal{PT}$)  symmetric Rosen-Morse potential. In this study, the stability of solution is examined by numerical analysis to show that solitons are stable for some parameter ranges. The dynamical properties such as evolution and transverse energy flow for both self-focusing and self-defocusing nonlinear mode are also analyzed.  The obtained results are useful for experimental designs and applications in related fields. 
\end{abstract}

\maketitle

\section{Introduction}
The introduction of $\mathcal{PT}$ operation has paved a new way into realm of undiscovered physical systems.  In recent years, nonlinear $\mathcal{PT}-$symmetric systems have become a fascinating area in research \cite{1}.  Both theoretical and experimental investigations are developing the subject into vast regimes.  The idea of $\mathcal{PT}-$symmetry and the possibility for investigation of real energy spectrum using this symmetry, despite the presence of complex terms in potential, was originated from Bender and Boettcher in 1998 \cite{2}. In the $\mathcal{PT}-$symmetry operation, the parity reflection ($\mathcal{P}$) is defined by $\hat{p}\rightarrow-\hat{p}$, $\hat{x}\rightarrow-\hat{x}$, $i\rightarrow i$ and time-reversal ($\mathcal{T}$) is defined as $\hat{p} \rightarrow -\hat{p}$, $\hat{x} \rightarrow \hat{x}$, $i\rightarrow-i$ \cite{3,4}. The experimental realization of such $\mathcal{PT}-$symmetric potential has led to discover new kind of optical lattices, lasers, metamaterials, etc \cite{5,6,7,8}. In nonlinear optical devices the complex refractive index constitutes the $\mathcal{PT}-$symmetric potential which balances the loss and gain of optical pulse in the medium.  $\mathcal{PT}-$symmetry breaking is also an important physical phenomenon because some of the desired properties are achieved at the point of symmetry breaking \cite{9,10} and in the symmetry broken phase \cite{10a,10b,10c,10d}.  $\mathcal{PT}-$symmetry breaking exploited in many experiments \cite{11,12,12a,12b}.

Many theoretical studies have been done on nonlinear Schr\"odinger equation (NLS) with $\mathcal{PT}-$symmetric potentials such as harmonic potential \cite{13}, sextic anharmonic double-well potential \cite{14}, Scarf \Romannum{2} potential \cite{15}, Rosen-Morse potential \cite{16} and Gaussian potential \cite{9,17}. More interestingly, in Ref. [16] the authors found that the $\mathcal{PT}-$symmetric Rosen-Morse potential with NLS equation does not accommodate stable soliton solutions for any potential parameters. This is due to the nonvanishing terms in the imaginary part of the potential. In this study, we extend the analysis of Rosen-Morse potential with CGL equation to attain stable soliton solutions. The CGL is one the important equation describing a nonlinear optical system \cite{18} and investigations on such systems are hardly done due to the non-invariant $\mathcal{PT}-$symmetry of the system. Additional parameters in CGL equation give us the freedom to create new stable soliton solution in this system. The potential is no more $\mathcal{PT}-$symmetric but near $\mathcal{PT}-$symmetric \cite{19} due to the presence of asymmetric terms. 

Through this work we reflect the dynamical properties of a near $\mathcal{PT}-$symmetric potential by playing with the system parameters. We present some novel results in which the near $\mathcal{PT}-$symmetric Rosen-Morse potential accommodates stable soliton solutions. We have extensively studied the stability of soliton solutions along with the dynamical properties such as evolution dynamics and energy flow. Our work is also extended to both self-focusing and self-defocusing nonlinear modes.

\section{A physical model of the system}
In a physical model of $\mathcal{PT}-$symmetric system, the loss and gain is balanced. But the physical system we consider has some additional physical parameters which lead to imbalance in gain and loss in the medium, thus leading to $\mathcal{PT}-$symmetric breaking. The transmission of such an optical lattice is given by CGL equation \cite{19,20,21} with complex constants and potentials as
\begin{align}\label{eq1}
\begin{split}
i\frac{\partial\Psi}{\partial z}+(\alpha_{1}+i \alpha_{2})\frac{\partial^{2}\Psi}{\partial x^{2}}+[V(x)+i W(x)] \Psi +\\ \sigma(\beta_{1}+i \beta_{2})|\Psi|^{2} \Psi = 0,
\end{split}
\end{align} 
where $\Psi(x,z)$ is the amplitude of complex electric field of the optical pulse, $z$ is the propagation distance, $x$ is the spatial coordinate, $\alpha_{1}$ is the diffraction coefficient, $\beta_{1}$ is the Kerr-nonlinearity coefficient, $\alpha_{2}$ spectral filtering, $\beta_{2}$ gives the nonlinear gain/loss in the system, $\sigma=\pm 1$ gives the self-focusing and self-defocusing nonlinear mode, $V(x)$ is the real part of potential and $W(x)$ is the imaginary part of potential.  For invariant $\mathcal{PT}-$symmetry the potential should be of the form, $V(-x)=V(x)$ and $W(-x)=-W(x)$. It is noted that the CGL equation is non$-\mathcal{PT}$ invariant due to the presence of complex coefficients. 

Our analysis starts by assuming stationary solution in the form $\Psi(x,z)=\phi(x)e^{i\mu z}$, where $\mu$ is the real valued propagation constant and $\phi(x)$ is the complex field.
\begin{figure*}[!ht]
\begin{subfigure}[b]{0.49\textwidth}
\includegraphics[width=1\textwidth]{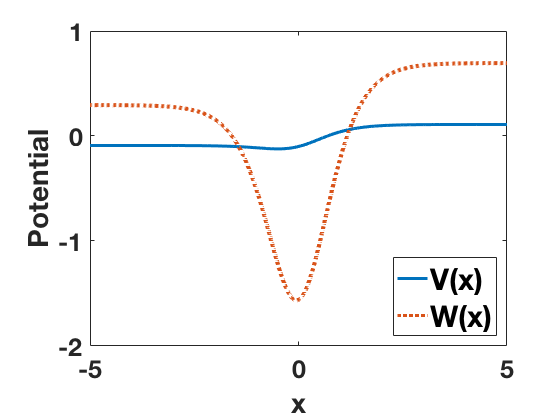}
\caption{}
\end{subfigure}
\begin{subfigure}[b]{0.49\textwidth}
\includegraphics[width=1\textwidth]{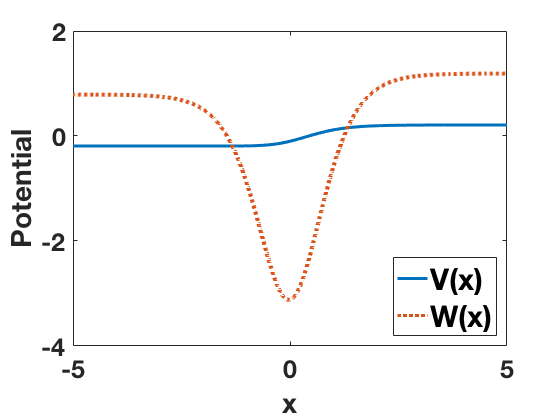}
\caption{}
\end{subfigure}
\begin{subfigure}[b]{0.49\textwidth}
\includegraphics[width=1\textwidth]{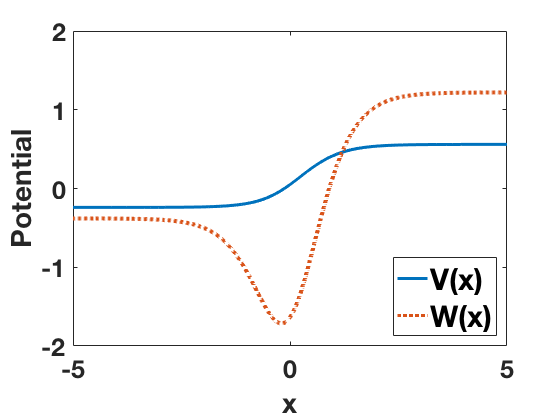}
\caption{}
\end{subfigure}
\begin{subfigure}[b]{0.49\textwidth}
\includegraphics[width=1\textwidth]{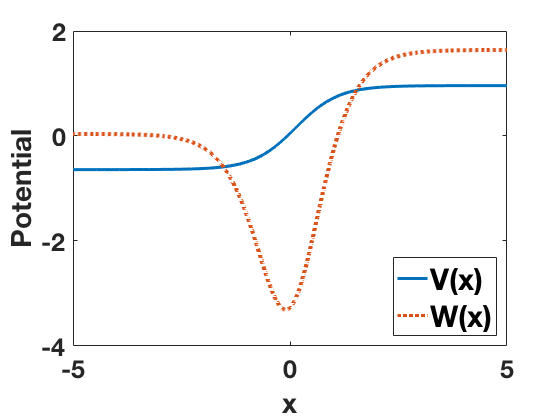}
\caption{}
\end{subfigure}
\caption{(Color online) The nature of near $\mathcal{PT}-$symmetric Rosen-Morse potential, both $V(x)$ and $W(x)$. \textbf{(a)} $b=0.1,\alpha_{2}=-0.5$ and $\beta_{2}=0.5$;  \textbf{(b)} $b=0.1,\alpha_{2}=-1$ and $\beta_{2}=1$; \textbf{(c)} $b=0.4,\alpha_{2}=-0.5$ and $\beta_{2}=0.5$; and \textbf{(d)} $b=0.4,\alpha_{2}=-1$ and $\beta_{2}=1$.}
\label{fig1}
\end{figure*}
Substituting this solution into Eq. (\ref{eq1}), we obtain the following second-order ordinary differential equation of the form
\begin{align}\label{eq2}
(\alpha_{1}+ i\alpha_{2}) \frac{d^{2}\phi}{dx^{2}} +[V(x)+iW(x)]\phi \nonumber  \\ +\sigma(\beta_{1}+i\beta_{2})|\phi|^{2}\phi-\mu \phi=0.
\end{align}
We use the condition $\mu=\alpha_{1}$ for the above equation. 

In order to investigate the optical beam propagation of  (\ref{eq1}), we consider the near $\mathcal{PT}-$symmetric Rosen-Morse potential which admits stationary soliton solution is of the form
\begin{subequations}\label{eq3}
\begin{align}
\label{eq3a} V(x)= & -a(a+1)\sech^{2}(x)-2\left(\frac{\alpha_2}{\alpha_1}\right)b\tanh(x)+\frac{b^2}{\alpha_1}, \\
\label{eq3b} W(x)=& 2b\tanh(x)+W_1\sech^2(x)+\alpha_2\left(\frac{b^2}{\alpha_1^2}-1\right),
\end{align}
\end{subequations}
where $W_{1}=2\alpha_{2}-(a(a+1)+2\alpha_{1})\left( \frac{\beta_{2}}{\beta_{1}}\right)$, $a$ and $b$ are positive real values for describing the strength of the potential. The original Rosen-Morse potential is given by
\begin{subequations}\label{eq4}
\begin{eqnarray}\label{eq4a}
V(x)=-c(c+1)\sech^{2}(x),
\end{eqnarray}
\begin{eqnarray}\label{eq4b}
W(x)=2d \tanh (x),
\end{eqnarray}
\end{subequations}
where $c$ and $d$ describe the strength of real and imaginary parts of the potential respectively. Comparing Eqs. (\ref{eq4}) with Eqs. (\ref{eq3}), it is evident that the existence of additional terms in the potential leads to near $\mathcal{PT}-$symmetry. Here  the physical model (\ref{eq1}) is described by many system parameters and for our convenience, we fix the values $a=0.1$ and $\alpha_{1}=\beta_{1}=1$. The existence of system parameters $\alpha_{2}$ and $\beta_{2}$ breaks the $\mathcal{PT}-$symmetry of the system. If the physical effects described by $\alpha_{2}$ and $\beta_{2}$ is very negligible then the system leads to $\mathcal{PT}-$symmetry.   In other words if the values of these parameters are zero then the potential is exactly $\mathcal{PT}-$symmetric. The qualitative nature of near $\mathcal{PT}$-symmetric Rosen-Morse potential for some parameter values are given in Fig. \ref{fig1}. Unlike the $\mathcal{PT}-$symmetric Rosen-Morse potential, here both $V(x)$ and $W(x)$ are non-vanishing asymptotically.

We have taken the values for parameters in such a way that we can illustrate the dependence of parameter in both potentials.  Initially we fix the value for $b$ in Figs. \ref{fig1}(a)-\ref{fig1}(b) and varied $\alpha_{2}$ and $\beta_{2}$. There is a appreciable shift in $W(x)$ and the depth of the potential $W(x)$ increases.  Then we change the value of $b$ for the same variations of $\alpha_{2}$ and $\beta_{2}$ as seen in Figs. \ref{fig1}(c)-\ref{fig1}(d).  The increase in value of $b$ is also contributing to the shape of the potential.  In addition, 
we also observe that the asymmetry of both the potentials increase with $b$.  These are the major factors that actually control the potential. In a physical system the values of $\alpha_{2}$ and $\beta_{2}$ are important because these values are externally controlled to make slight modifications to the potentials. In other way we can say that these parameters can be used to tune the system to reach a particular regime where we can achieve stable propagation of light pulses.

\section{Stationary soliton solution and linear stability analysis}
\subsection{Self-focusing nonlinear mode}
Our investigation of stationary solution starts with self-focusing nonlinearity, $\sigma=1$. Now the system admits an exact solution for Eq. (\ref{eq2}) in the form
\begin{eqnarray}\label{eq5}
\phi(x)=\sqrt{\frac{a(a+1)+2\alpha_{1}}{\beta_{1}}} \sech(x) e^{\frac{ibx}{\alpha_{1}}}	
\end{eqnarray}
For this particular solution, we impose the condition $\mu=\alpha_{1}$ for which the above is true.
\begin{figure}[h!]
\begin{subfigure}[b]{0.23\textwidth}
\includegraphics[width=1\textwidth]{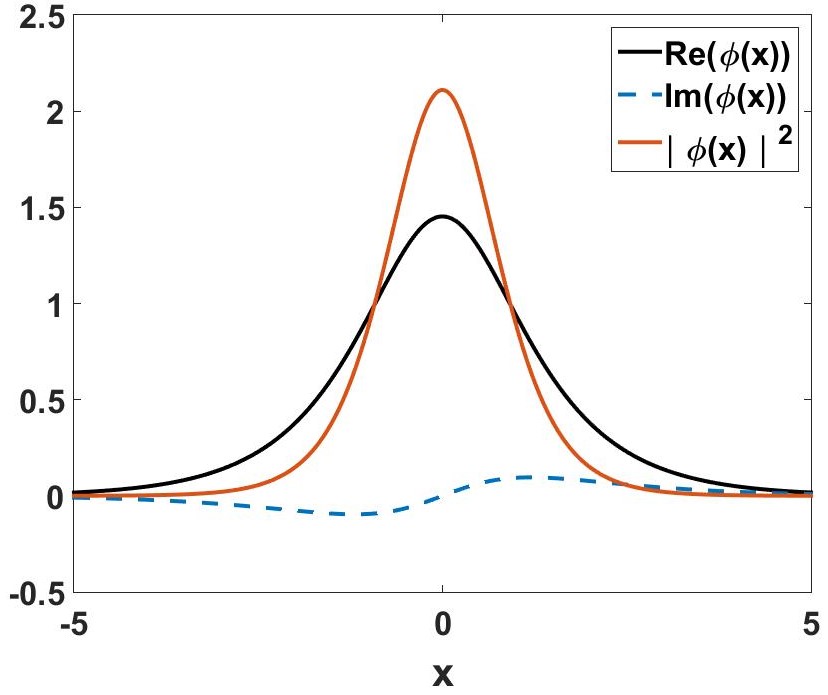}
\caption{}
\end{subfigure}
\begin{subfigure}[b]{0.23\textwidth}
\includegraphics[width=1\textwidth]{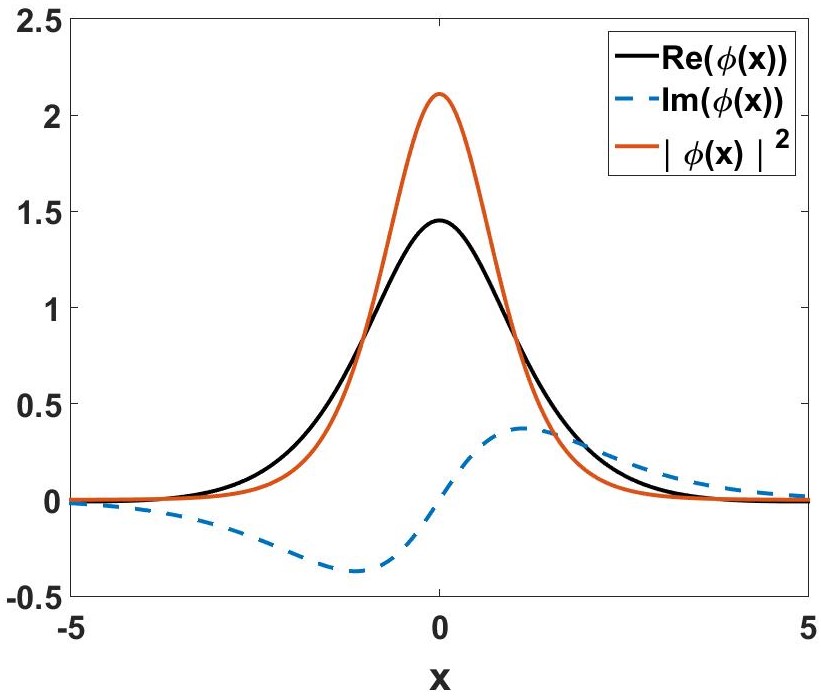}
\caption{}
\end{subfigure}
\caption{(Color online) Plot for stationary solution (\ref{eq5}) showing real part, imaginary part and $|\phi(x)|^2$ for \textbf{(a)} $ b=0.1$ and \textbf{(b)} $b=0.4.$}
\label{fig2}
\end{figure}
The profile of stationary soliton solution of Eq. (\ref{eq2}) is shown in Fig. \ref{fig2}. When we increase the parameter $b$ from $0.1$ to $0.4$, there is no change in total intensity of stationary solution.  But we can observe an increase in amplitude of imaginary part and decrease in the width of real part of stationary solution. 

The stability of this solution is evaluated by adding infinitesimal perturbation with the stationary solution. If such a perturbation leads to deviation from the original solution, then the solution is unstable. The perturbed solution \cite{22} is given by
\begin{eqnarray}
\label{eq6}
\Psi(x,z) = (\phi(x)+\epsilon [f(x)e^{\lambda z}+g^{*}(x)e^{\lambda^{*} z}])e^{i\mu z},
\end{eqnarray}
where $\epsilon \ll 1$, $f(x)$ and $g(x)$ are very small perturbation functions and $\lambda$ is the stability parameter. Substituting the perturbed solution in Eq.  (\ref{eq1}) and linearizing with respect to $\epsilon$, one can obtain a set of equations for $f(x)$ and $g(x)$ which can be solved by forming a matrix of the system of equations. 

\begin{figure*}[ht!]
\begin{subfigure}[b]{0.23\textwidth}
\includegraphics[width=1\textwidth]{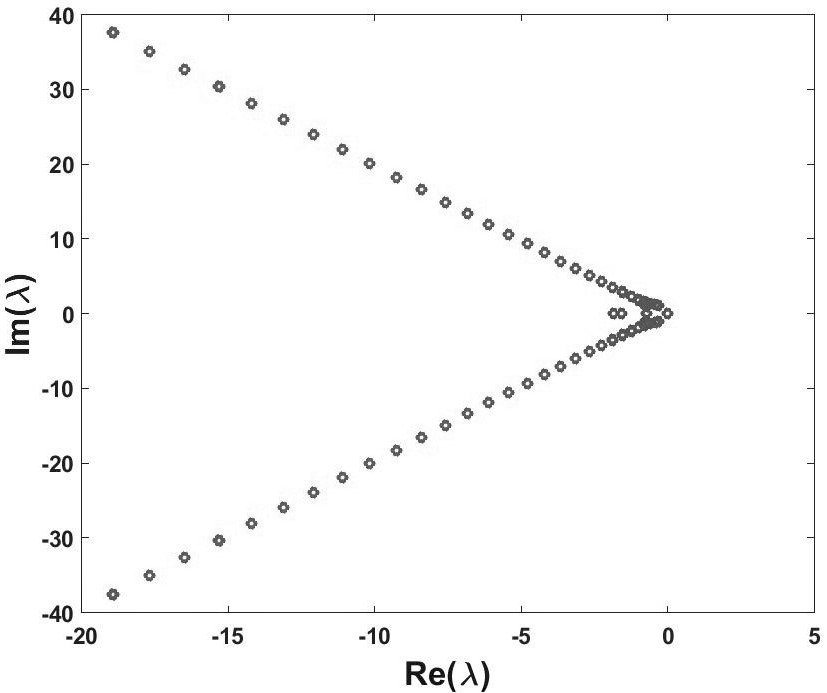}
\caption{}
\end{subfigure}
\begin{subfigure}[b]{0.23\textwidth}
\includegraphics[width=1\textwidth]{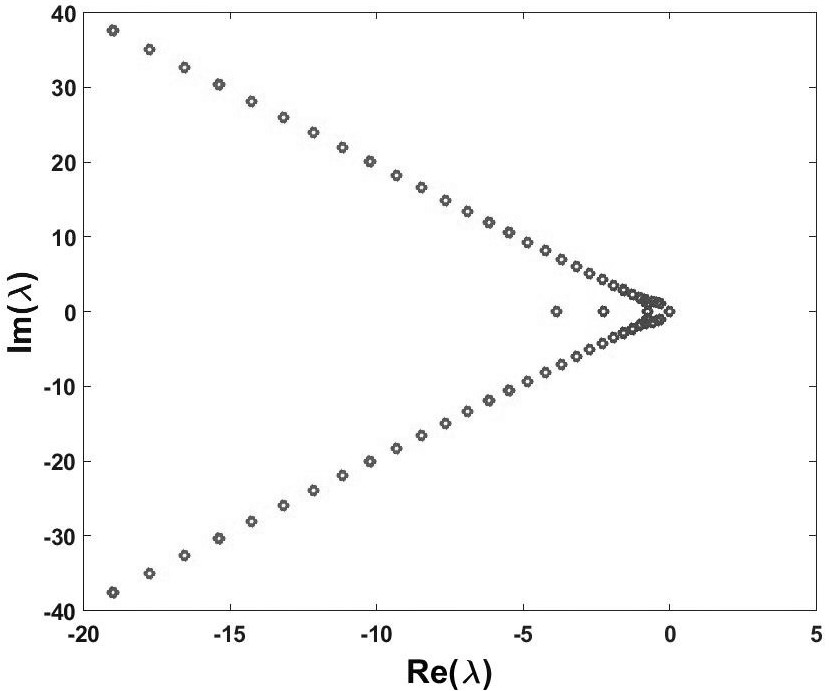}
\caption{}
\end{subfigure}
\begin{subfigure}[b]{0.23\textwidth}
\includegraphics[width=1\textwidth]{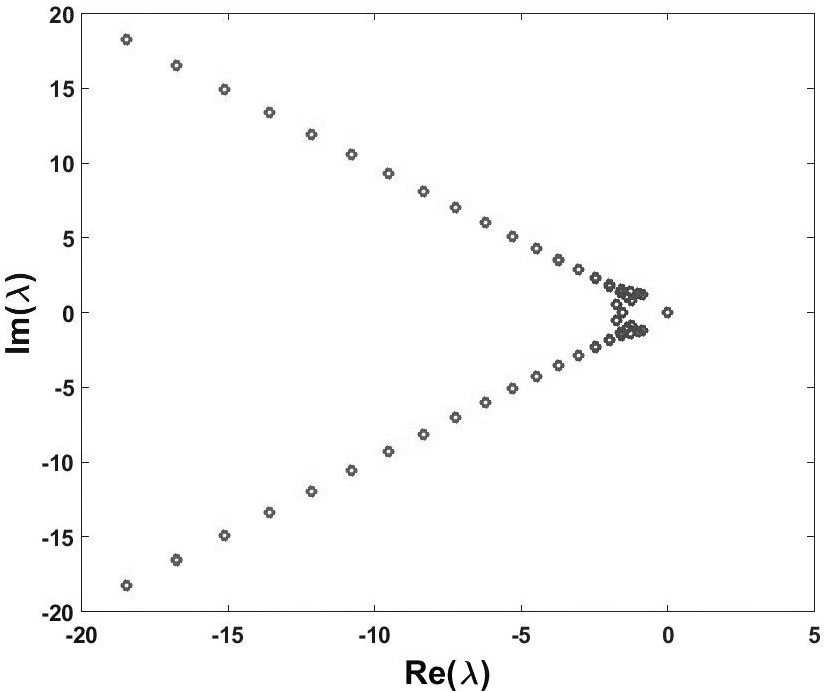}
\caption{}
\end{subfigure}
\begin{subfigure}[b]{0.23\textwidth}
\includegraphics[width=1\textwidth]{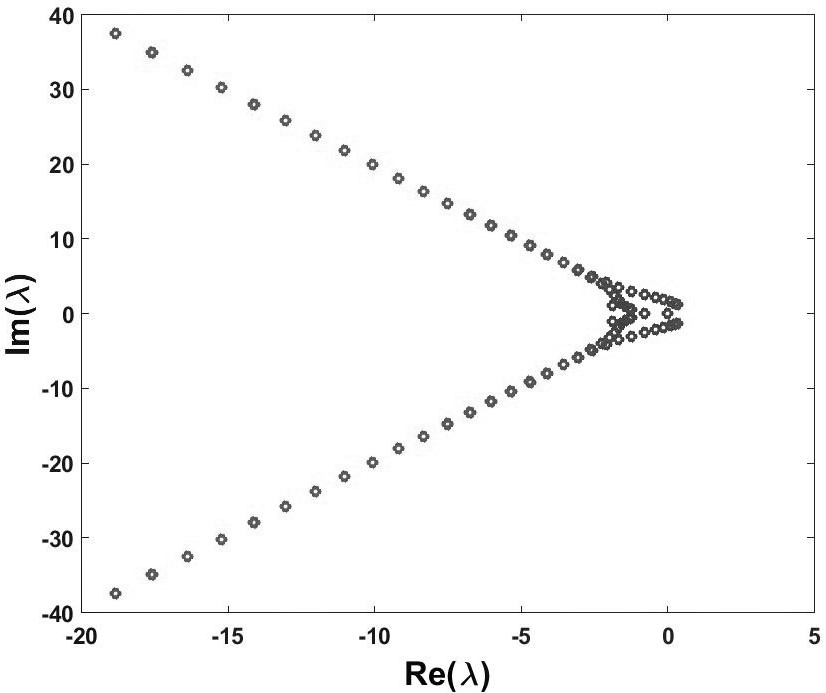}
\caption{}
\end{subfigure}
\begin{subfigure}[b]{0.23\textwidth}
\includegraphics[width=1\textwidth]{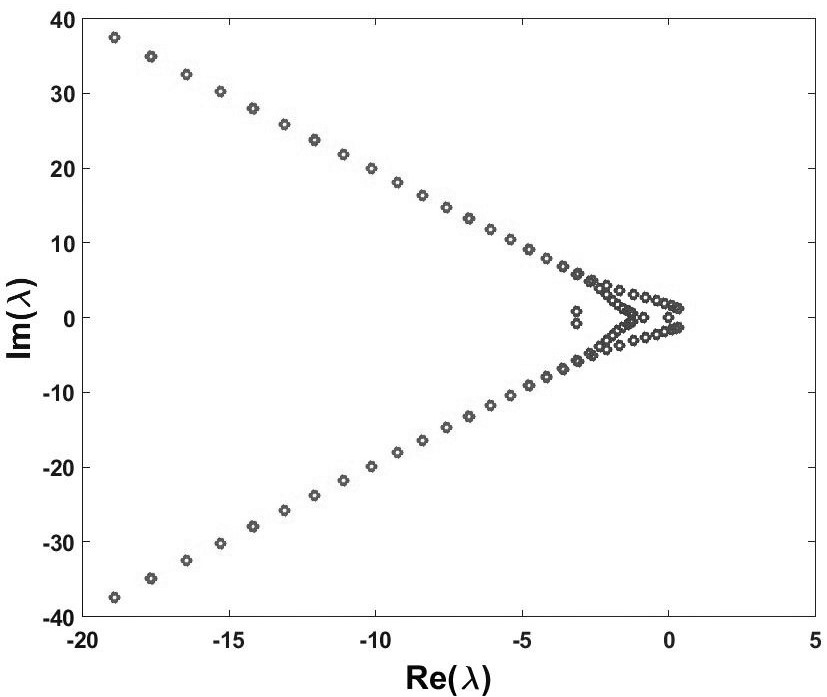}
\caption{}
\end{subfigure}
\begin{subfigure}[b]{0.23\textwidth}
\includegraphics[width=1\textwidth]{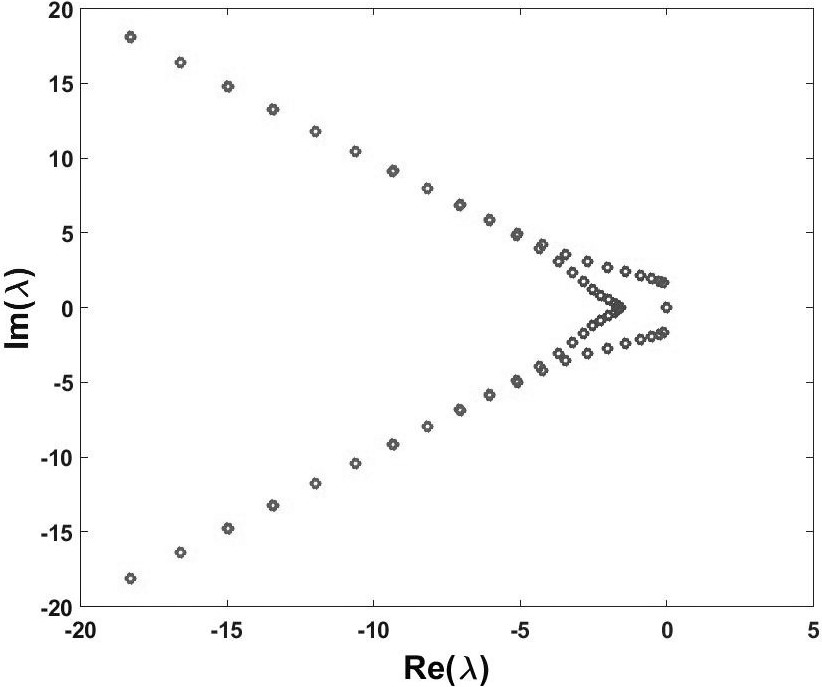}
\caption{}
\end{subfigure}
\begin{subfigure}[b]{0.23\textwidth}
\includegraphics[width=1\textwidth]{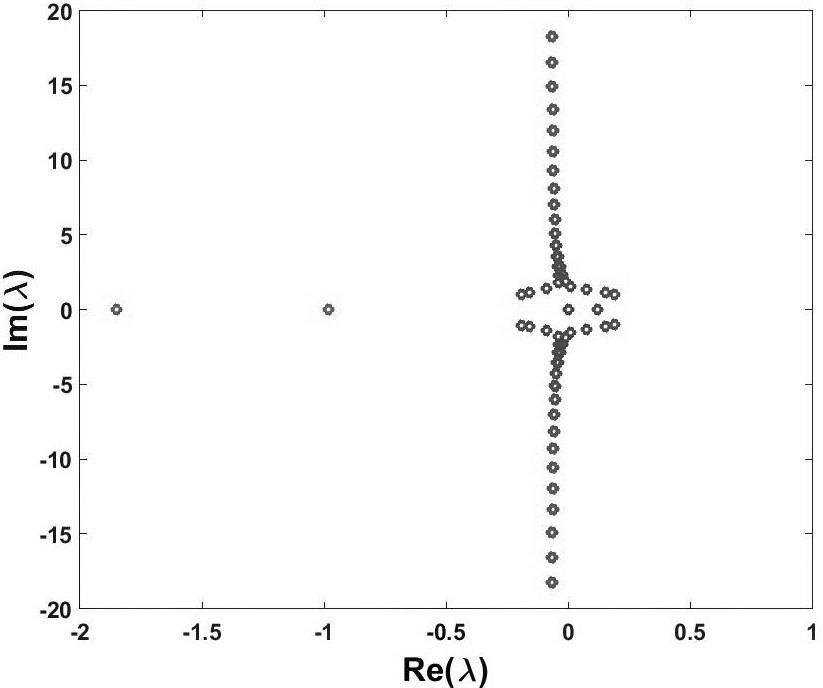}
\caption{}
\end{subfigure}
\begin{subfigure}[b]{0.23\textwidth}
\includegraphics[width=1\textwidth]{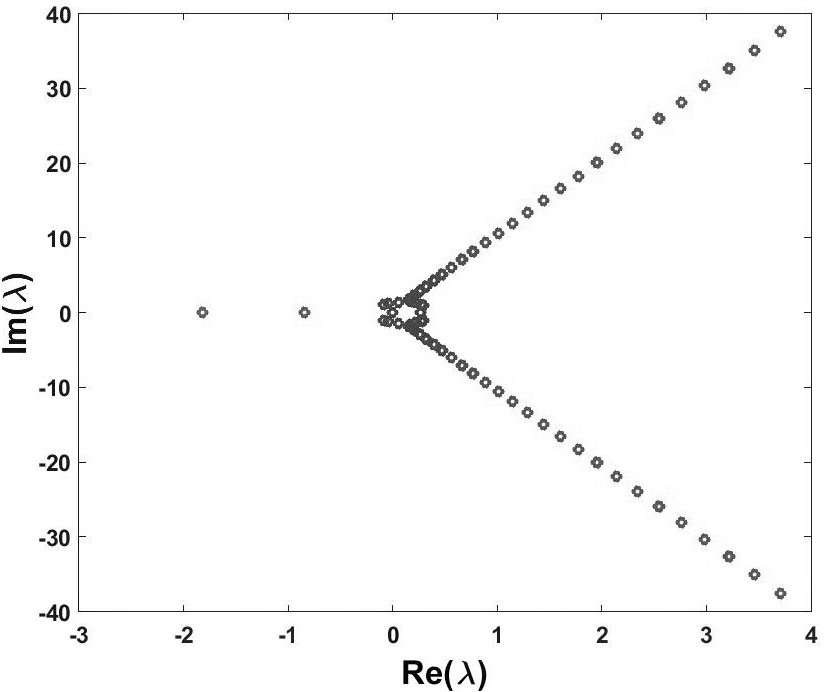}
\caption{}
\end{subfigure}
\caption{The eigenvalue spectrum of stability analysis for (\ref{eq1}) with \textbf{(a)} $b=0.1$, $\alpha_{2}=-0.5$, $\beta_{2}=0.5$; \textbf{(b)} $b=0.1$, $\alpha_{2}=-0.5$, $\beta_{2}=1$; \textbf{(c)} $b=0.1$, $\alpha_{2}=-1$, $\beta_{2}=0.5$; \textbf{(d)} $b=0.4$, $\alpha_{2}=-0.5$, $\beta_{2}=0.5$; \textbf{(e)} $b=0.4$, $\alpha_{2}=-0.5$, $\beta_{2}=1$; \textbf{(f)} $b=0.4$, $\alpha_{2}=-1$, $\beta_{2}=0.5$; \textbf{(g)} $b=0.1$, $\alpha_{2}=0$, $\beta_{2}=0.5$; and \textbf{(h)} $b=0.1$, $\alpha_{2}=1$, $\beta_{2}=0.5$.}
\label{fig3}
\end{figure*}

The linear eigenvalue problem is described by 
\begin{gather}
\begin{pmatrix} 
L_{0} &  L_{1}\\
 -L_{1}^{*} & -L_{0}^{*} 
\end{pmatrix}  
\begin{pmatrix}
f(x)\\
g(x)
\end{pmatrix}=-i\lambda
\begin{pmatrix}
f(x)\\
g(x)
\end{pmatrix}
\end{gather}
where $L_{0}=(\alpha_{1}+i\alpha_{2})\partial_{xx}+V(x)+iW(x)+2(\beta_{1}+i\beta_{2})|\phi|^{2}-\mu$ and $L_{1}=(\beta_{1}+i\beta_{2})\phi^{2}$.
This eigenvalue equation can be solved using Fourier collocation method \cite{23}.  Stable solutions are only present if the eigenvalues have a negative real part or purely imaginary. This means that when the system is perturbed from a singular point, the perturbation will eventually decay for negative real part of the eigenvalues. The stationary solution with purely imaginary eigenvalues oscillates around the singular point for small perturbation.  In case of positive eigenvalues, the terms in perturbation diverge and stability cannot be achieved even for infinitesimal perturbation. Figure \ref{fig3} illustrates the eigenvalues for different parameter values.

\begin{figure*}[!ht]
\begin{subfigure}[b]{0.23\textwidth}
\includegraphics[width=1\textwidth]{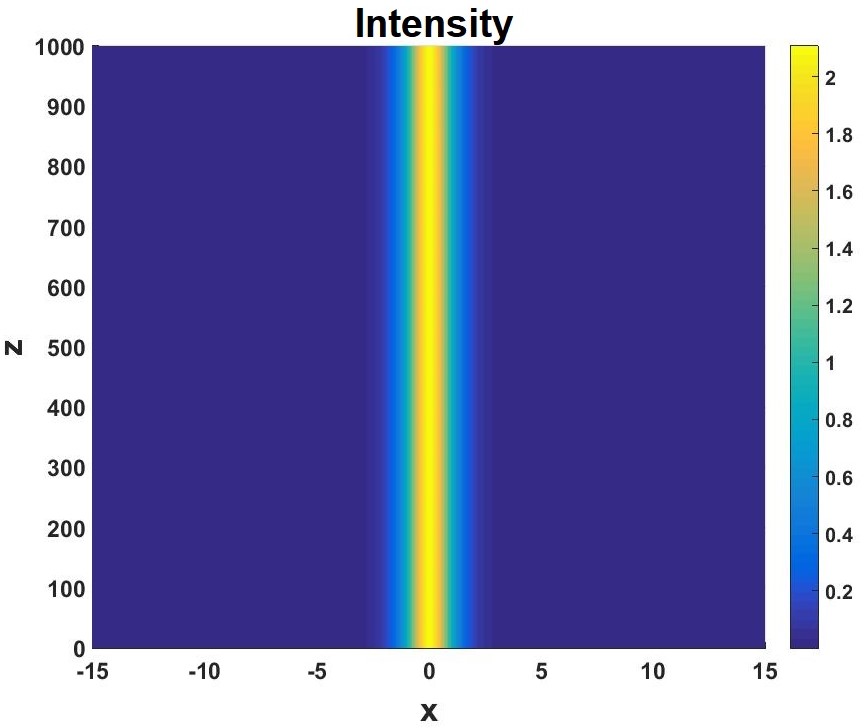}
\caption{}
\end{subfigure}
\begin{subfigure}[b]{0.23\textwidth}
\includegraphics[width=1\textwidth]{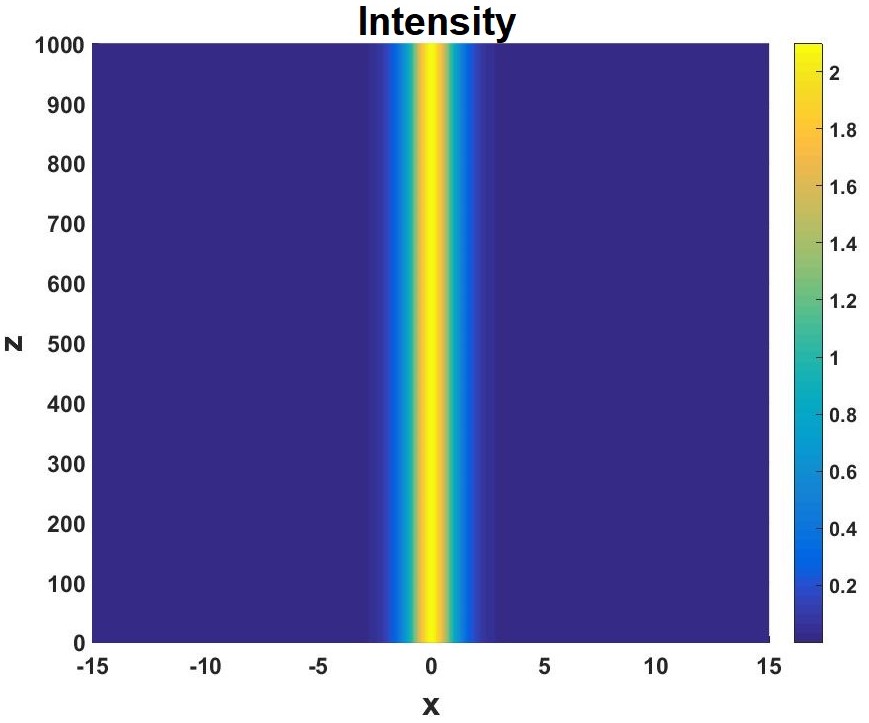}
\caption{}
\end{subfigure}
\begin{subfigure}[b]{0.23\textwidth}
\includegraphics[width=1\textwidth]{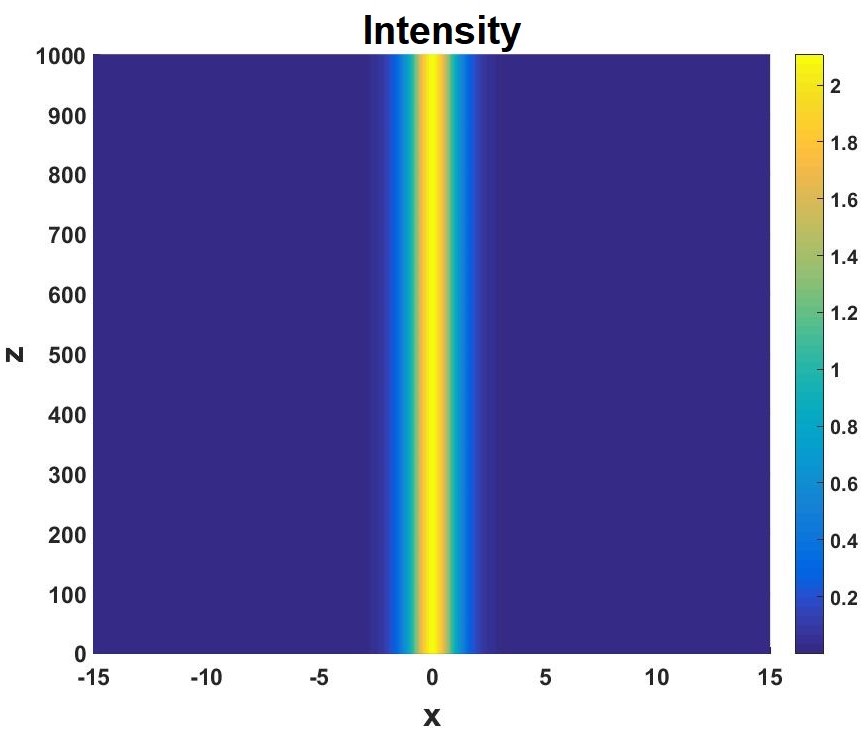}
\caption{}
\end{subfigure}
\begin{subfigure}[b]{0.23\textwidth}
\includegraphics[width=1\textwidth]{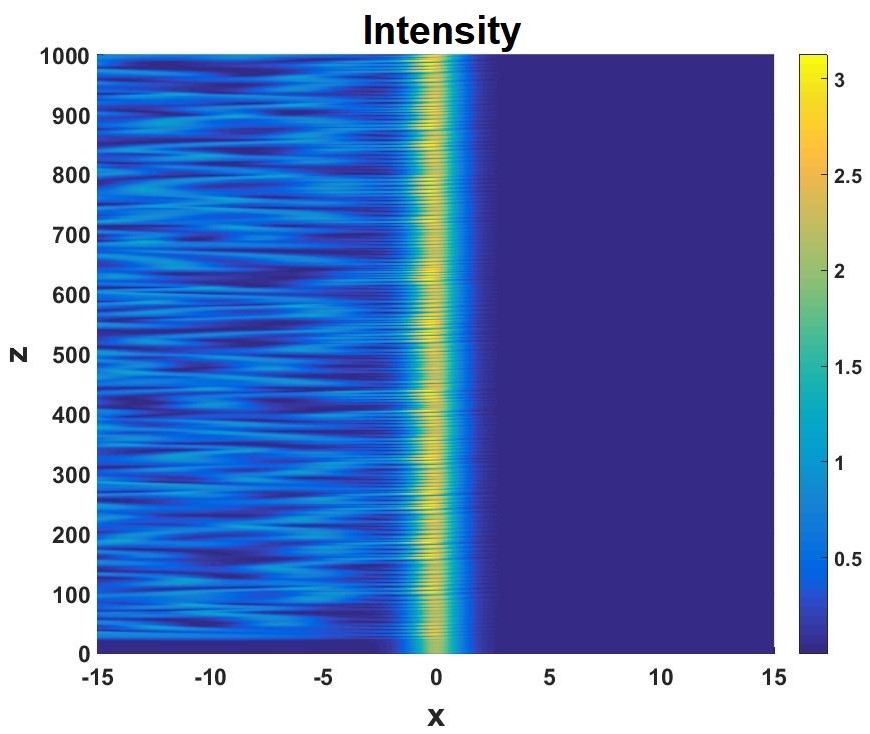}
\caption{}
\end{subfigure}
\begin{subfigure}[b]{0.23\textwidth}
\includegraphics[width=1\textwidth]{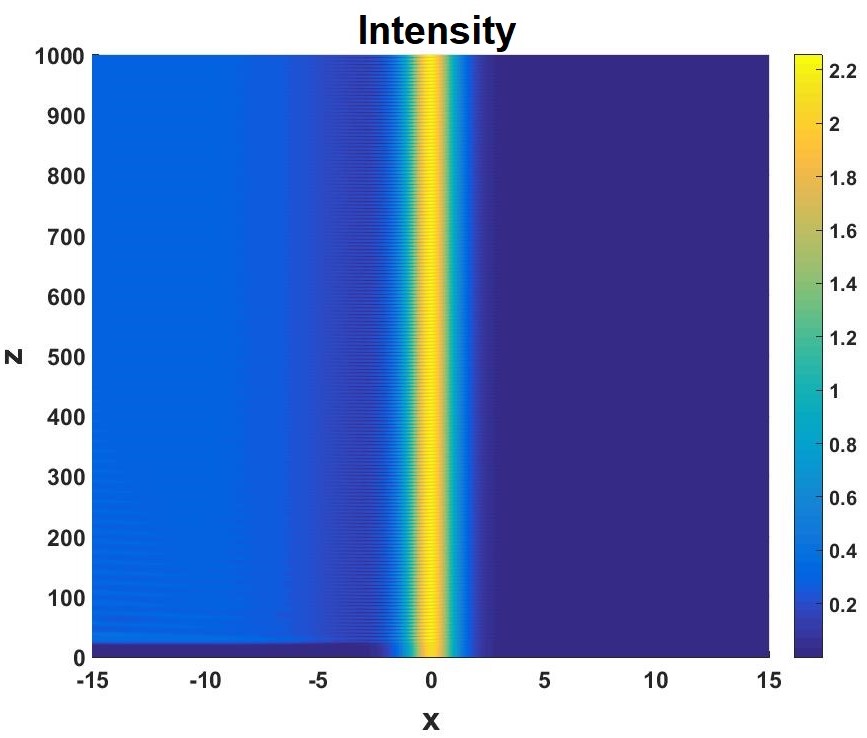}
\caption{}
\end{subfigure}
\begin{subfigure}[b]{0.23\textwidth}
\includegraphics[width=1\textwidth]{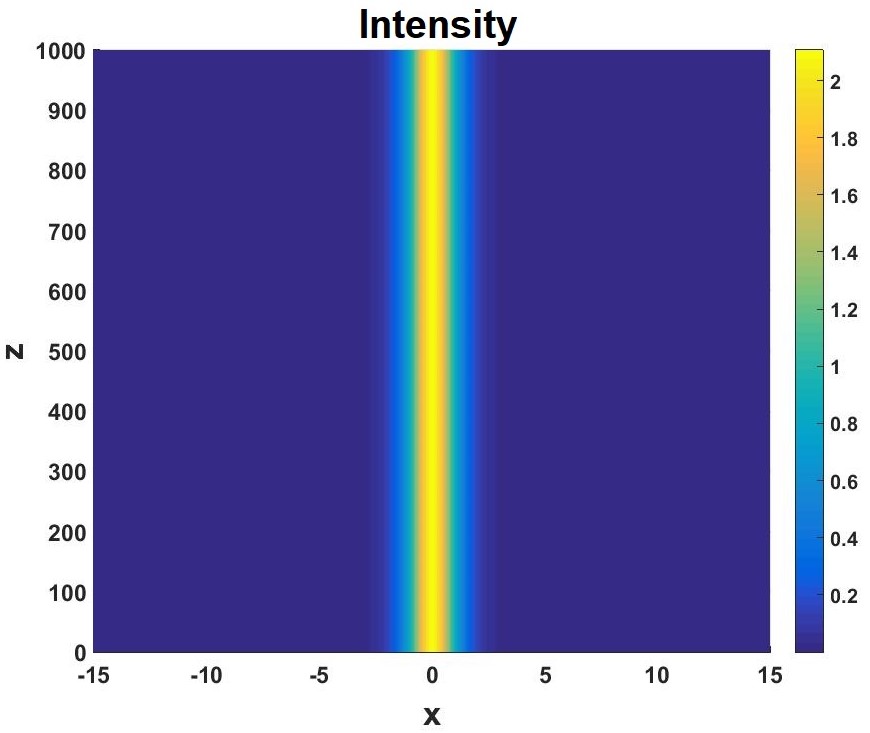}
\caption{}
\end{subfigure}
\begin{subfigure}[b]{0.23\textwidth}
\includegraphics[width=1\textwidth]{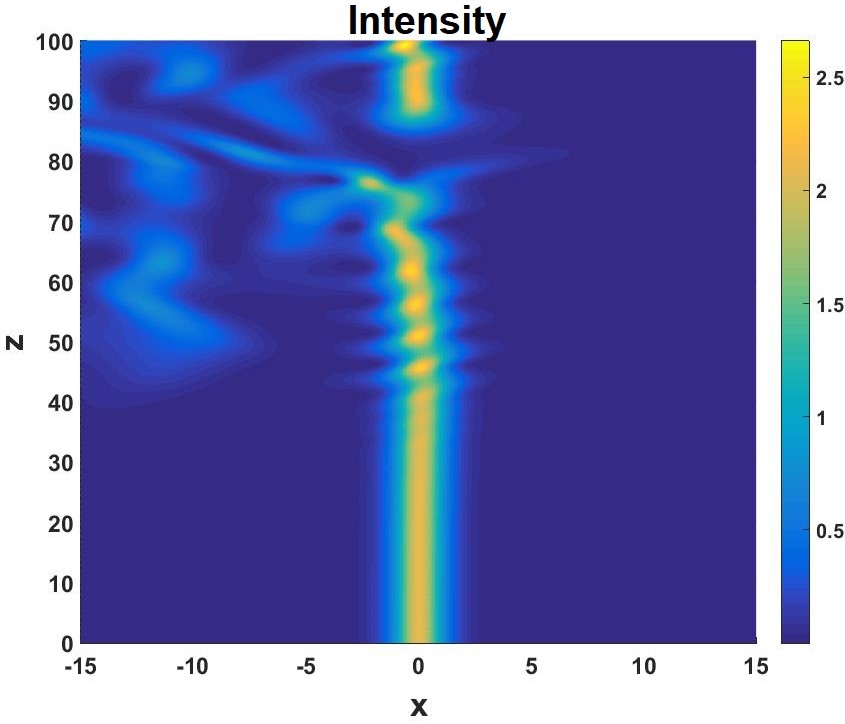}
\caption{}
\end{subfigure}
\begin{subfigure}[b]{0.23\textwidth}
\includegraphics[width=1\textwidth]{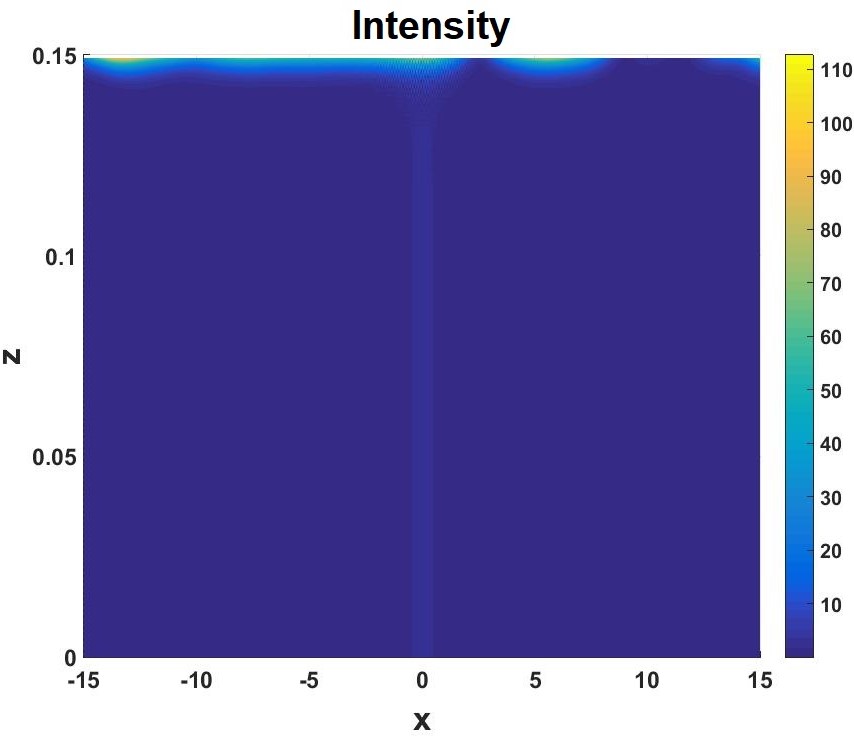}
\caption{}
\end{subfigure}
\caption{(Color online) Intensity plot for stable evolution of soliton along the $z$ direction for same values of Fig. \ref{fig3}.}
\label{fig4}
\end{figure*}

Thus we have obtained the parameter values for stable and unstable modes using linear stability analysis.  The stable regions are small ranges in different parameters.   All soliton solutions for our considered system are stable only for $\alpha_{2}<0$.  Values for stable solutions are obtained and eigenvalues for the same are given in Figs. \ref{fig3}(a)-\ref{fig3}(f).  Figures \ref{fig3}(g) and \ref{fig3}(h) show eigenvalues for unstable solutions.  Here, the eigenvalues have positive real part which will not converge when perturbed.

We follow the convention that the soliton propagates in $z -$ direction.  Figure \ref{fig4} represents intensity plot of solitons for the same parameter values as in Fig. \ref{fig3}. It is interesting to note from Figs. \ref{fig4}(d) and \ref{fig4}(e) where the soliton profile is not smooth as other stable solutions. During propagation, the amplitude oscillates rather than maintaining a constant value.  Thus we infer that there exists some cases where the amplitude of solitons oscillates during propagation without affecting their stability. These oscillations are unknown from stability analysis. For stable soliton solutions, the maximum intensity is at $x=0$ plane and reduces to zero when $x\rightarrow\pm5$. In Fig. \ref{fig4}(g) the soliton is unstable, deviates from its path and dissipates. But in Fig. \ref{fig4}(h) as $\alpha_{2}$ increases, the soliton is not able to propagate even a little distance, where it scatters due to $\alpha_2 > 0$. Thus, we find that due to the $\mathcal{PT}-$symmetry breaking, the near $\mathcal{PT}-$symmetric Rosen-Morse potential accommodates stable soliton solutions. 

\subsection{Self-defocusing nonlinear mode}
Extending our insight into self-defocusing nonlinearity, for which $\sigma=-1$, the exact soliton solution for Eq. (\ref{eq2}) is given by 	
\begin{eqnarray}\label{eq8}
\phi(x)=\sqrt{-\frac{a(a+1)+2\alpha_{1}}{\beta_{1}}} \text{sech}(x) e^{\frac{ibx}{\alpha_{1}}},
\end{eqnarray}
where all the terms are already discussed in Sec. \Romannum{2}.  The difference between the solutions of focusing and defocusing mode is the only change in the sign of the term inside square root of the amplitude of soliton. Here also the condition for obtaining this solution is $\mu=\alpha_{1}$.
\begin{figure}[h!]
\begin{subfigure}[b]{0.23\textwidth}
\includegraphics[width=1\textwidth]{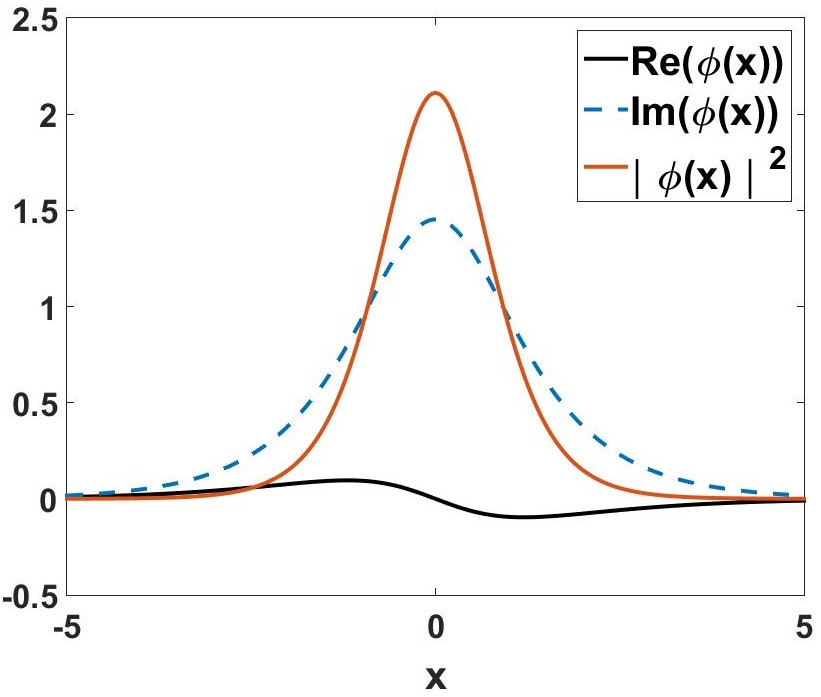}
\caption{}
\end{subfigure}
\begin{subfigure}[b]{0.23\textwidth}
\includegraphics[width=1\textwidth]{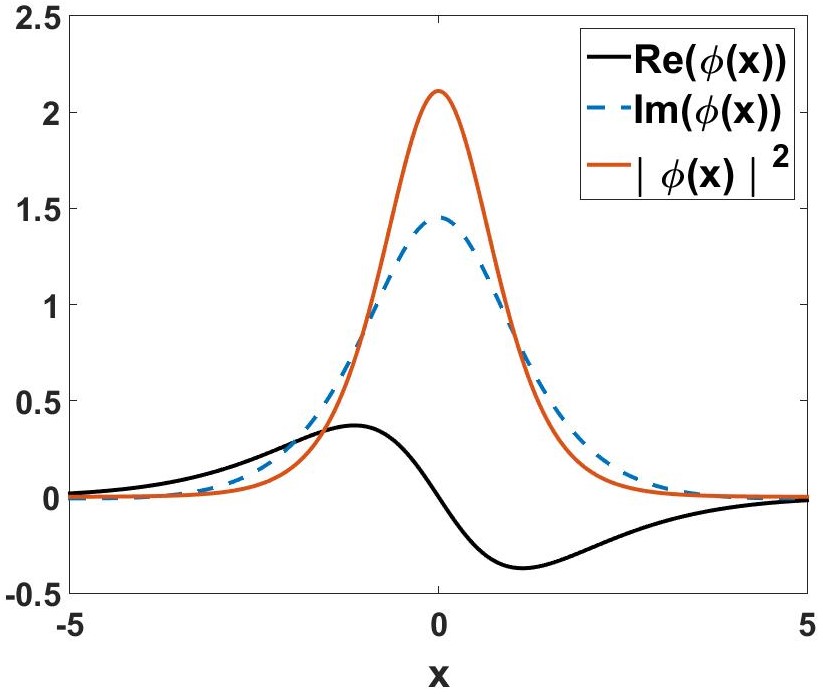}
\caption{}
\end{subfigure}
\caption{(Color online) Plot for stationary solution (\ref{eq8}) showing real part, imaginary part and $|\phi(x)|^2$ for \textbf{(a)} $b=0.1$ and \textbf{(b)} $b=0.4$.}
\label{fig5}
\end{figure}

\begin{figure}[h!]
\begin{subfigure}[b]{0.23\textwidth}
\includegraphics[width=1\textwidth]{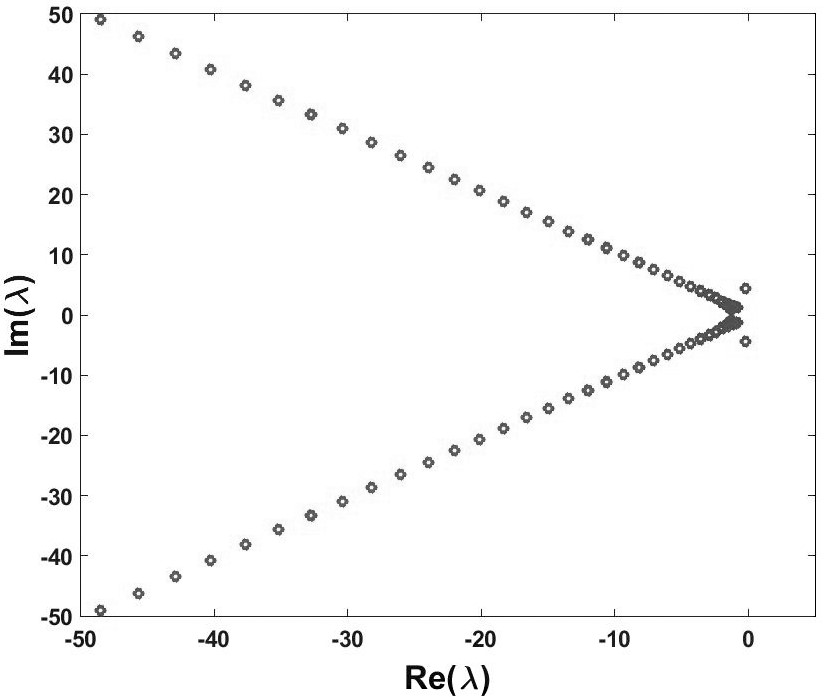}
\caption{}
\end{subfigure}
\begin{subfigure}[b]{0.23\textwidth}
\includegraphics[width=1\textwidth]{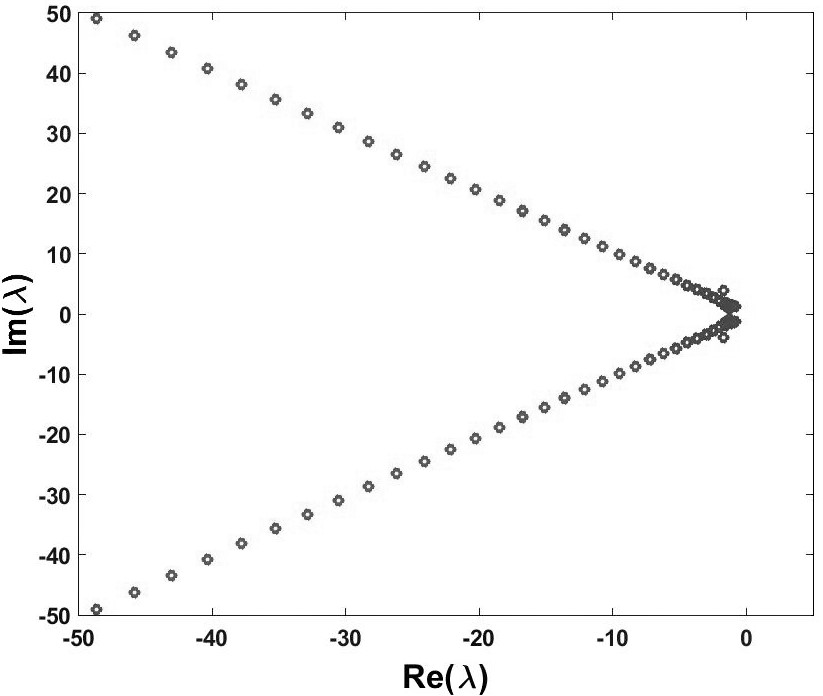}
\caption{}
\end{subfigure}
\begin{subfigure}[b]{0.23\textwidth}
\includegraphics[width=1\textwidth]{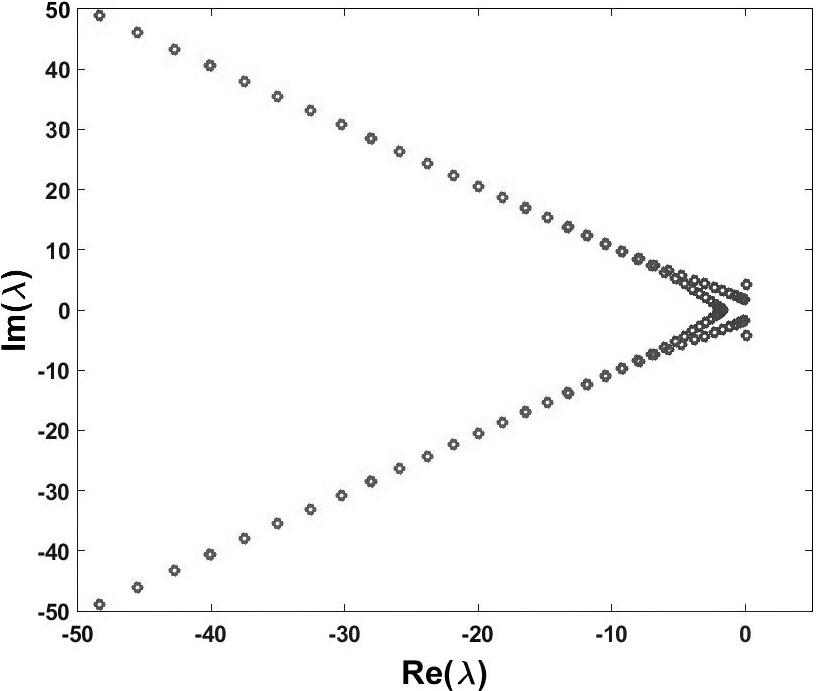}
\caption{}
\end{subfigure}
\begin{subfigure}[b]{0.23\textwidth}
\includegraphics[width=1\textwidth]{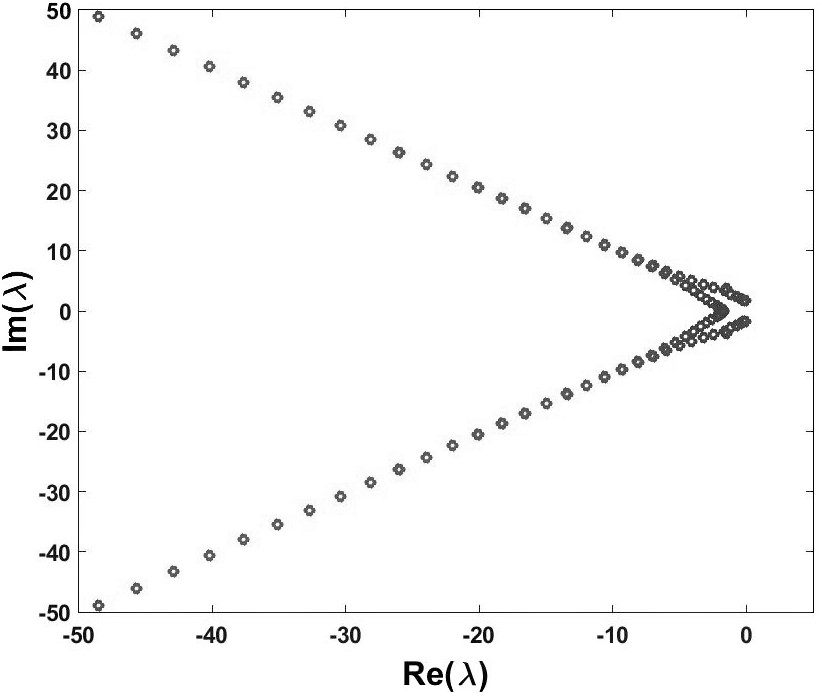}
\caption{}
\end{subfigure}
\caption{The eigenvalue spectrum of stability analysis for defocusing nonlinear mode with  \textbf{(a)} $b=0.1$, $\beta_{2 }=0.15$;  \textbf{(b)} $b=0.1$, $\beta_{2}=-0.15$;  \textbf{(c)} $b=0.4$, $\beta_{2 }=0.15$;  and  \textbf{(d)} $b=0.4$, $\beta_{2}=-0.15$.}
\label{fig6}
\end{figure}

The solution, Eq. (\ref{eq8}) is given by Fig. \ref{fig5} with the real part, imaginary part and amplitude of the exact soliton solution.  In Fig. \ref{fig5}(a), we can see that the imaginary part of solution is more dominant than the real part due to the negative term inside the square root (see Eq.  (\ref{eq8})).  When we increase the parameter $b$ from $0.1$ to $0.4$, there is no change in total intensity of stationary solution as shown in Fig. \ref{fig5}(b).  In addition, we notice the increase in amplitude of real part and decrease in the width of imaginary part of stationary solution.  The linear stability analysis for the above solution is studied as mentioned earlier and our results are plotted in Fig. \ref{fig6}  for $\alpha_{2}=-1$.  

From Fig. \ref{fig6} the stability analysis for self-defocusing nonlinear mode also shows the same sign pattern of $\alpha_{2}$, but with less number of combinations. The range of values are very short compared with self-focusing nonlinearity. The evolution of stationary solution for the parameters obtained by stability analysis confirms the smoothness of soliton. 

\begin{figure}[h!]
\begin{subfigure}[b]{0.23\textwidth}
\includegraphics[width=1\textwidth]{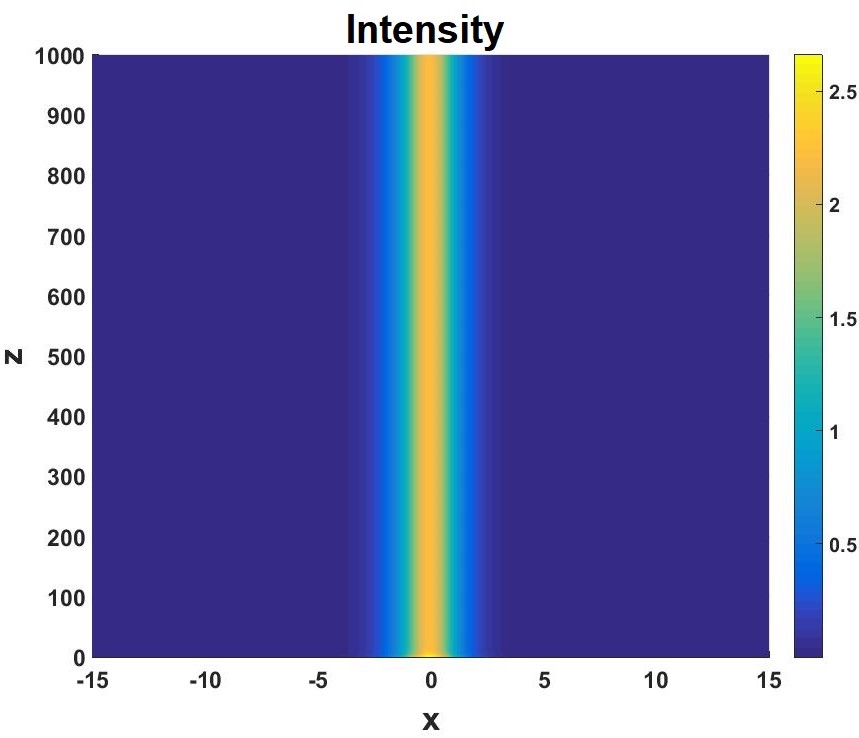}
\caption{}
\end{subfigure}
\begin{subfigure}[b]{0.23\textwidth}
\includegraphics[width=1\textwidth]{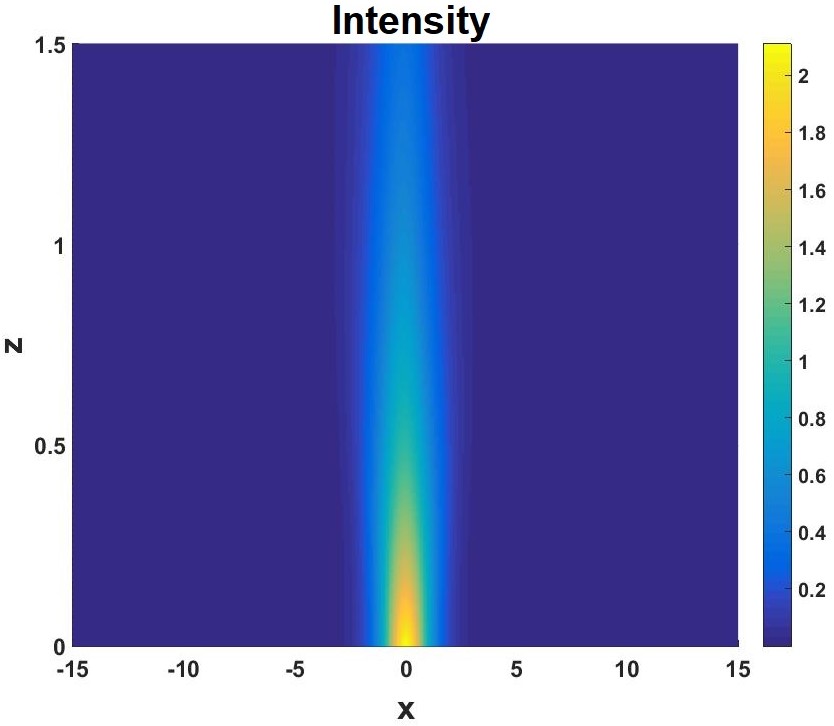}
\caption{}
\end{subfigure}
\begin{subfigure}[b]{0.23\textwidth}
\includegraphics[width=1\textwidth]{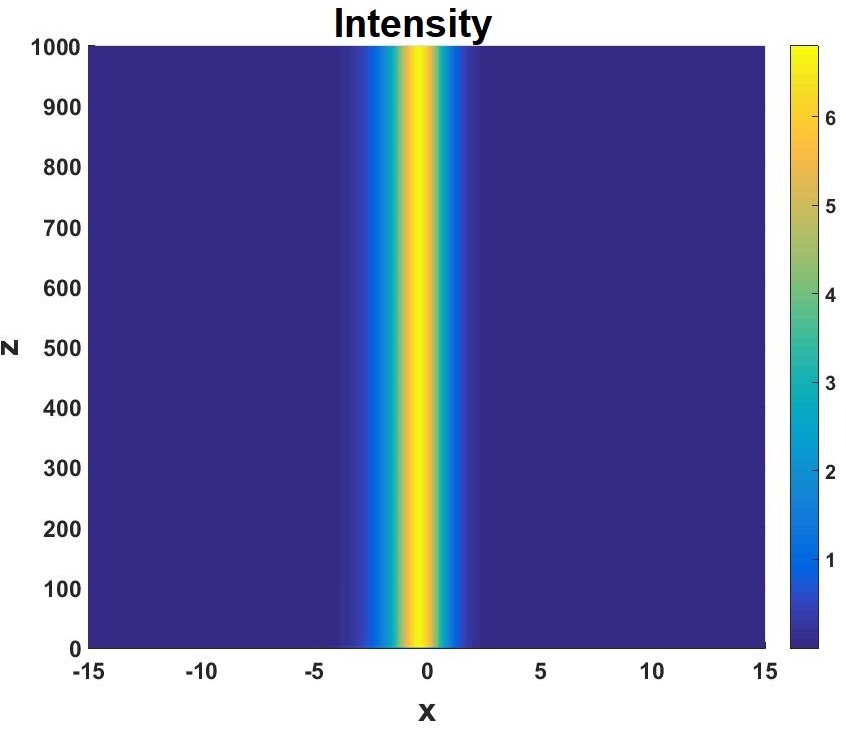}
\caption{}
\end{subfigure}
\begin{subfigure}[b]{0.23\textwidth}
\includegraphics[width=1\textwidth]{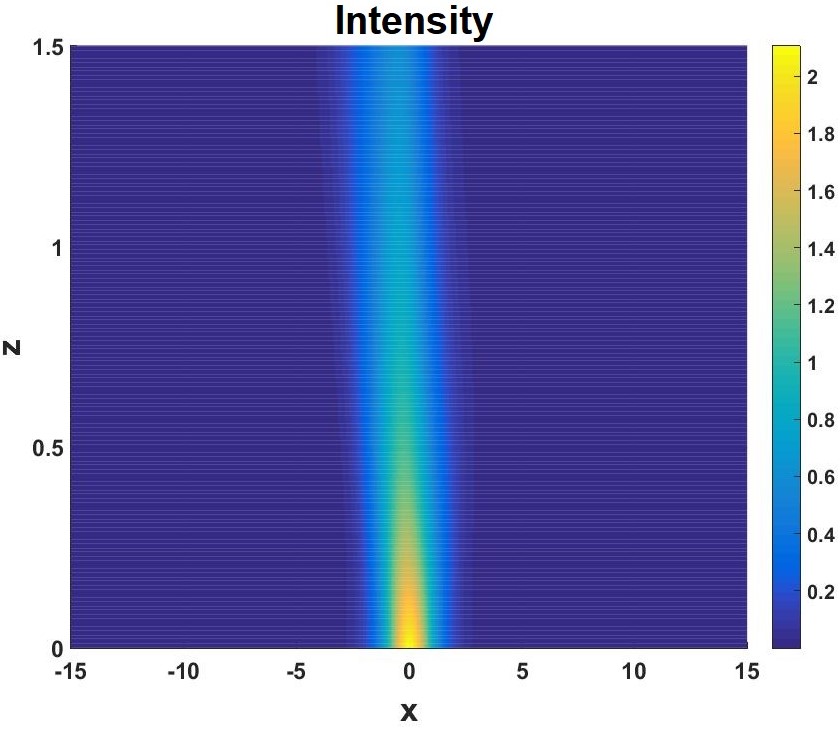}
\caption{}
\end{subfigure}
\caption{(Color online) Intensity plot for stable evolution of soliton for same values as in Fig. \ref{fig6}.}
\label{fig7}
\end{figure}

The evolution of soliton is given in Fig. \ref{fig7}.  Even the solutions are stable for the given values of parameters, the soliton profile does not guarantee propagation without loss is intensity. Figs. \ref{fig7}(a) and \ref{fig7}(c) show the evolution of stable soliton solution with a small variation in amplitude in the beginning of propagation. When we reduce the value of $\beta_{2}$, the soliton dissipates after propagating a small distance as in Figs. \ref{fig7}(b) and \ref{fig7}(d). Comparing with the self-focusing nonlinear mode, for the stable soliton solutions, self-defocusing nonlinear mode has narrow range for parameters. Keeping other parameters constant and increasing the value of $b$ the intensity of soliton increases which means we can amplify the intensity of light pulse with modulating the potential strength, that is the refractive index of wave guide.

\section{Energy flow for exact soliton solution}
\subsection{Self-focusing nonlinear mode}
\begin{figure}[h!]
\begin{subfigure}[b]{0.23\textwidth}
\includegraphics[width=1\textwidth]{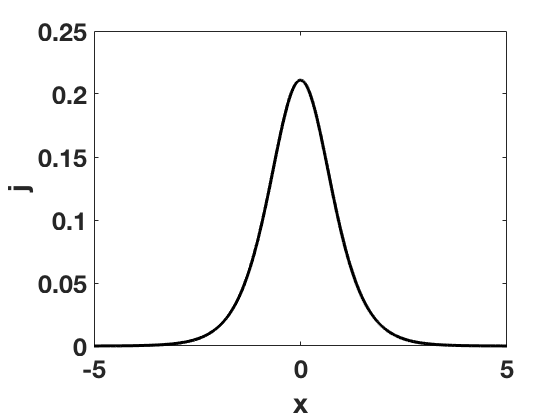}
\caption{}
\end{subfigure}
\begin{subfigure}[b]{0.23\textwidth}
\includegraphics[width=1\textwidth]{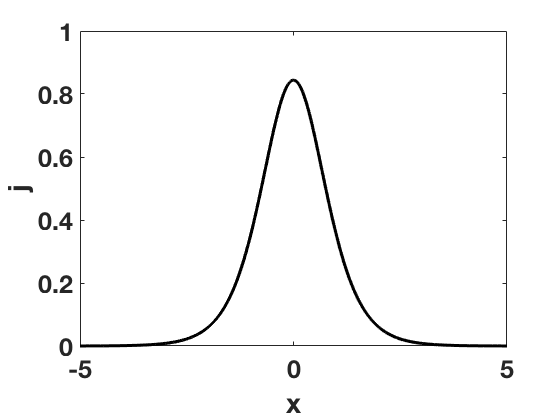}
\caption{}
\end{subfigure}
\caption{Energy flux $(j)$ with respect to spatial coordinate for self-focusing mode:   \textbf{(a)} $b=0.1$ and \textbf{(b)} $b=0.4$.}
\label{fig8}
\end{figure}

\begin{figure}[h!]
\begin{subfigure}[b]{0.23\textwidth}
\includegraphics[width=1\textwidth]{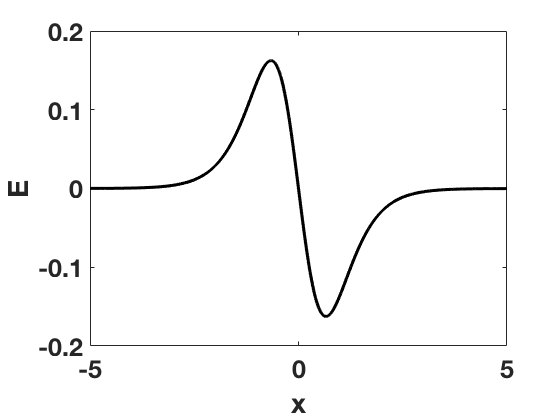}
\caption{}
\end{subfigure}
\begin{subfigure}[b]{0.23\textwidth}
\includegraphics[width=1\textwidth]{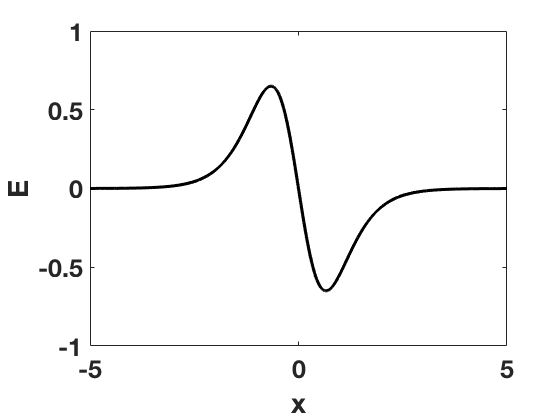}
\caption{}
\end{subfigure}
\caption{The gain or loss distribution of energy $(E)$ with respect to spatial coordinate for \textbf{(a)} $b=0.1$ and \textbf{(b)} $b=0.4$.}
\label{fig9}
\end{figure}

The equation of continuity for energy flow (where the energy is not conserved) of CGL Eq. (\ref{eq1}) gives us the relation between the energy density ($\rho = |\Psi|^2$) and the energy flux $(j)$ which is given by \cite{19} 
\begin{eqnarray}\label{eq9}
E = \frac{\partial\rho}{\partial z}+\frac{\partial j}{\partial x},
\end{eqnarray}
where $E$ determines the gain or loss distribution of energy and the energy flux is described by $j=\frac{i}{2}(\phi\phi_{x}^{*}-\phi_{x}\phi^{*})$.  If the system is conservative then $E=0$ which means the loss and gain in the system is balanced. When the loss and gain of a system is not balanced, energy will flow from one region to another region.  For the system of our interest with $\sigma=1$, we find
\begin{align}\label{eq10}
E= & \frac{-2b}{\alpha_{1} \beta_{1}}(a+a^{2}+2\alpha_{1})\text{sech}^{2}(x)\text{tanh}(x), \\
j= & \frac{b(a(a+1)+2\alpha_1)}{\alpha_1\beta_1}\sech^2(x).
\end{align}
Figure \ref{fig8} shows the variation of energy flux with spatial coordinate. 
The plot is similar to the shape of the stationary soliton, that is the value of $j$ peaks at $x=0$, which implies that maximum intensity is transferred through $x=0$. The energy flux depends on the value of $b$ since $a, \alpha_{1}$  and $\beta_{1}$ are fixed. The direction and quantity of energy flow are illustrated in Fig. \ref{fig9}.   Here we observe that the light pulse gains energy in $x<0$ region
and looses energy in $x>0$ region.  In other words, the energy flow is effected from left to right.  An important observation from Eq. (\ref{eq10}) is that the energy flow due to unbalanced loss and gain will not depend on $\alpha_{2}$ and $\beta_{2}$. Thus, the strength of the potential controls the energy flow in near $\mathcal{PT}-$symmetric potentials.
\subsection{Self-defocusing nonlinear mode}
\begin{figure}[h!]
\begin{subfigure}[b]{0.23\textwidth}
\includegraphics[width=1\textwidth]{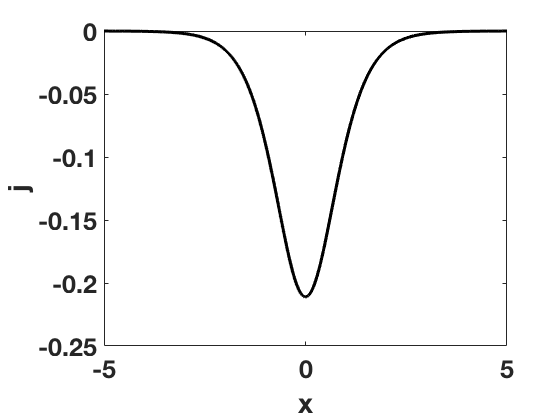}
\caption{}
\end{subfigure}
\begin{subfigure}[b]{0.23\textwidth}
\includegraphics[width=1\textwidth]{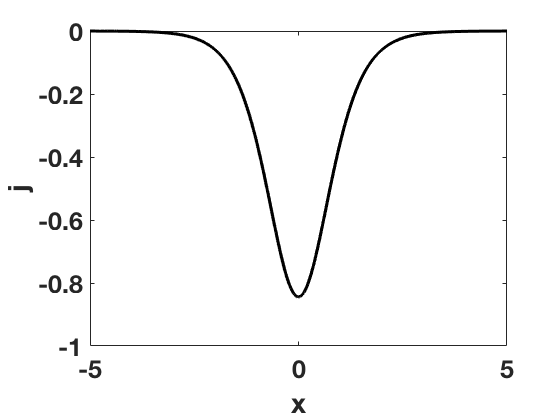}
\caption{}
\end{subfigure}
\caption{Variation of energy flux $(j)$ with respect to spatial coordinate for self-defocusing mode: \textbf{(a)} $b=0.1$ and \textbf{(b)} $b=0.4$.}
\label{fig10}
\end{figure}

\begin{figure}[h!]
\begin{subfigure}[h!]{0.23\textwidth}
\includegraphics[width=1\textwidth]{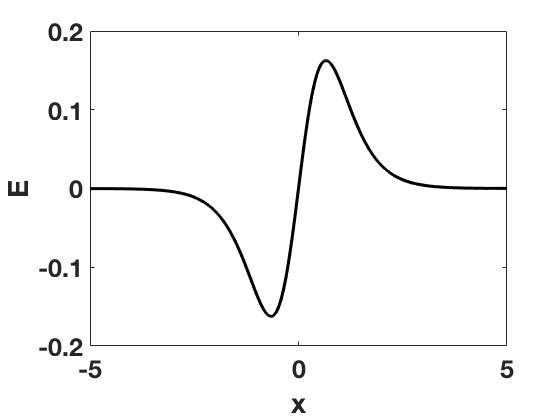}
\caption{}
\end{subfigure}
\begin{subfigure}[h!]{0.23\textwidth}
\includegraphics[width=1\textwidth]{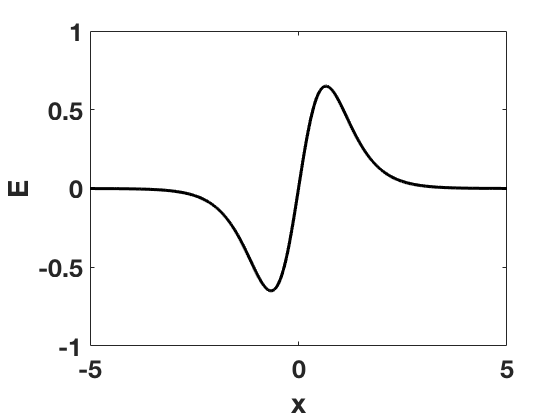}
\caption{}
\end{subfigure}
\caption{The gain or loss distribution of energy $(E)$ with respect to spatial coordinate for \textbf{(a)} $b=0.1$ and \textbf{(b)} $b=0.4$.}
\label{fig11}
\end{figure}

Similarly, for the self-defocusing case $(\sigma=-1)$, the gain or loss distribution of energy and the energy flux are given by 
\begin{align}\label{eq11}
E=&\frac{2b}{\alpha_{1} \beta_{1}}(a+a^{2}+2\alpha_{1})\text{sech}^{2}(x)\text{tanh}(x),\\
j= & -\frac{b(a(a+1)+2\alpha_1)}{\alpha_1\beta_1}\sech^2(x).
\end{align}
The energy flux for the self-defocusing nonlinear mode is shown in Fig. \ref{fig10}.
Further, maximum energy flux is at $x=0$  and decreases on either side as earlier.  Figure \ref{fig11} gives the gain or loss distribution of energy in near $\mathcal{PT}-$symmetric unbalance in potential. It is clear that the gain or loss distribution of energy in negative $x-$axis corresponds to loss, and positive in $x-$axis corresponds to gain.  Altogether, for self-defocusing nonlinearity the distribution of energy flows from positive $x$ direction to negative $x$ direction. 
Here also the energy flow is independent of $\alpha_{2}$ and $\beta_{2}$ but dependent on the strength $(b)$ of the potential. Therefore, when the sign of nonlinearity in the CGL equation changes the direction of energy flow also changes.
\\
\section{Conclusion}
In our study, we modified the $\mathcal{PT}-$symmetric Rosen-Morse potential with NLS equation to a near $\mathcal{PT}-$symmetric Rosen-Morse potential with CGL equation.  Exact $\mathcal{PT}-$symmetric soliton solutions for this potential along with stability analysis and evolution of soliton solution for different parameter ranges are analyzed for both self-focusing and self-defocusing modes.  For some parameter ranges, the $\mathcal{PT}-$symmetric soliton in near $\mathcal{PT}-$symmetric potential shows excellent behavior compared to the $\mathcal{PT}-$symmetric potential where such solutions are unstable.  The stable soliton solutions are apparently possible for combination of $\alpha_{2}<0$ and $\beta_{2}\geq 0$.  We have also investigated the energy flow in the potential and observed that the direction of energy flow in self-focusing and self-defocusing nonlinear modes are opposite.  $\mathcal{PT}-$symmetry breaking is an important phenomenon to create near $\mathcal{PT}-$symmetric potentials, and analyzing the possibilities of stable soliton solutions can be further developed into experimental realizations in future.

\section*{Acknowledgments}
KM wishes to thank the Department of Science and Technology - Science and Engineering Research Board, Government of India, for providing National Post Doctoral fellowship under the Grant No. PDF/2016/001620/PMS.


\begin{thebibliography}{50}

\bibitem{1}
V. V. Konotop, J. Yang and D. A. Zezyulin, Rev. Mod. Phys. {\bf 88}, 035002 (2016).

\bibitem{2}
C. M. Bender and S. Boettcher, Phys. Rev. Lett. {\bf 80}, 5243 (1998).

\bibitem{3}
C. M. Bender, Contemp. Phys. {\bf 46}, 277 (2005).

\bibitem{4}
C. M. Bender, Rep. Prog. Phys. {\bf 70}, 947 (2007).

\bibitem{5}
A. Regensburger, C. Bersch, M. A. Miri, G. Onishchukov, D. N. Christodoulides and U. Peschel, Nature {\bf 488}, 167, (2012).

\bibitem{6}
H. Hodaei, M. A. Miri, M. Heinrich, D. N. Christodoulides and M. Khajavikhan, Science {\bf 346}, 975 (2014).

\bibitem{7}
D. R. Smith, J. B. Pendry and M. C. Wiltshire, Science {\bf 305}, 788 (2004).

\bibitem{8}
N. Lazarides and G. P. Tsironis, Phys. Rev. Lett. {\bf 110}, 053901 (2013).

\bibitem{9}
V. Achilleos, P. G. Kevrekidis, D. J. Frantzeskakis and R. Carretero-Gonz\'alez, Phys. Rev. A {\bf 86}, 013808 (2012).

\bibitem{10}
J. Yang, Opt. Lett. {\bf 39}, 5547 (2014).

\bibitem{10a}
H. Ramezani, T. Kottos, R. El-Ganainy and D. N. Christodoulides, Phys. Rev. A {\bf 82}, 043803 (2010).

\bibitem{10b}
T. Kottos, Nat. Phy. {\bf 6}, 166 (2010).

\bibitem{10c} 
S. Karthiga, V. K. Chandrasekar, M. Senthilvelan and M. Lakshmanan, Phys. Rev. A {\bf 94}, 023829 (2016).

\bibitem{10d}
S. Karthiga, V. K. Chandrasekar, M. Senthilvelan and M. Lakshmanan, Phys. Rev. A {\bf 95}, 033829 (2017).

\bibitem{11}
M. Lawrence, N. Xu, X. Zhang, L. Cong, J. Han, W. Zhang and S. Zhang,  Phys. Rev. Lett. {\bf 113}, 093901 (2014).

\bibitem{12}
L. Feng, Z. J. Wong, R. M. Ma, Y. Wang and X. Zhang, Science {\bf 346}, 972 (2014).

\bibitem{12a}
C.E. R\"{u}ter, K. G. Makris, R. El-Ganainy, D. N. Christodoulides, M. Segev and D. Kip, Nat. Phy. {\bf 6}, 192 (2010).

\bibitem{12b}
A. Regensburger, C. Bersch, M. A. Miri, G. Onishchukov, D. N. Christodoulides and U. Peschel, Nature {\bf 488}, 167 (2012).

\bibitem{13}
D. A. Zezyulin and V. V. Konotop, Phys. Rev. A {\bf 85}, 043840 (2012).

\bibitem{14}
Z. C. Wen and Z. Yan, Phys. Lett. A {\bf 379}, 2025 (2015).

\bibitem{15}
Z. H. Musslimani, K. G. Makris, R. El-Ganainy and D. N. Christodoulides, Phys Rev. Lett. {\bf 100}, 030402 (2008).

\bibitem{16}
B. Midya and R. Roychoudhury, Phys. Rev. A {\bf 87}, 045803 (2013).

\bibitem{17}
S. Hu, X. Ma, D. Lu, Z. Yang, Y. Zheng and W. Hu, Phys. Rev. A {\bf 84}, 043818 (2011).

\bibitem{18}
P. A. B\'elanger, L. Gagnon and C. Par\'e, Opt. Lett. {\bf 14}, 943 (1989).

\bibitem{19}
Y. Chen, Z. Yan and W. Liu, Optics Express, {\bf 26}, 33022 (2018).

\bibitem{20}
Y. He and D. Mihalache, JOSA B, {\bf 29}, 2554 (2012).

\bibitem{21}
Y. He and D. Mihalache, Phys. Rev. A {\bf 87}, 013812 (2013).

\bibitem{22}
E. A. Kuznetsov, A. M. Rubenchik and V. E. Zakharov, Phys. Rep. {\bf 142}, 103 (1986).

\bibitem{23}
J. Yang, Nonlinear Waves in Integrable and Non-Integrable Systems (Society for Industrial and Applied Mathematics, Philadelphia, PA, 2010).

\end{thebibliography}
\end{document}